\documentclass[runningheads]{llncs}
\usepackage{amsfonts}
\usepackage{amsmath}
\usepackage{amssymb}
\usepackage{enumerate}
\usepackage[shortlabels]{enumitem}
\usepackage[T1]{fontenc}
\usepackage{xspace}

\usepackage{hyperref}
\hypersetup{
     colorlinks=true,
     linkcolor=blue,
     filecolor=blue,
     citecolor = blue,
     urlcolor=blue,
     }

\usepackage{graphicx}
\usepackage{color}

\begin{document}
\title{On the Formalization of the Heat Conduction Problem in HOL}
\titlerunning{On the Formalization of the Heat Conduction Problem in HOL}
%
\author{Elif Deniz\inst{1} \and
Adnan Rashid\inst{2} \and
Osman Hasan\inst{2} \and Sofi\`{e}ne Tahar\inst{1}}
\authorrunning{E. Deniz et al.}
%
\institute{Department of Electrical and Computer Engineering\\
Concordia University, Montreal, Quebec, Canada\\
\email{\{e\_deniz,tahar\}@ece.concordia.ca}\\
 \and
{School of Electrical Engineering and Computer Science} \\
{National University of Sciences and Technology, Islamabad, Pakistan}\\
\email{\{adnan.rashid,osman.hasan\}@seecs.nust.edu.pk}}
\maketitle             

\begin{abstract}
Partial Differential Equations (PDEs) are widely used for modeling the physical phenomena and analyzing the dynamical behavior of many engineering and physical systems. The heat equation is one of the most well-known PDEs that captures the temperature distribution and diffusion of heat within a body. Due to the wider utility of these equations in various safety-critical applications, such as thermal protection systems, a formal analysis of the heat transfer is of utmost importance. In this paper, we propose to use higher-order-logic (HOL) theorem proving for formally analyzing the heat conduction problem in rectangular coordinates. In particular, we formally model the heat transfer as a one-dimensional heat equation for a rectangular slab using the multivariable calculus theories of the HOL Light theorem prover. This requires the formalization of the heat operator and formal verification of its various properties, such as linearity and scaling. Moreover, we use the separation of variables method for formally verifying the solution of the PDEs, which allows modeling the heat transfer in the slab under various initial and boundary conditions using HOL Light.

\end{abstract}
\section{Introduction}
Partial Differential Equations (PDEs) \cite{PDE_book} are commonly used for the mathematical formulation of the physical behavior of many engineering and physical systems. They capture the continuous dynamics of a system by providing a mathematical relationship between various components of the underlying system by incorporating changes in their associated properties. Due to these distinguishing features, they are broadly used in analyzing many physical phenomena such as, heat or sound propagation, electrodynamics, quantum mechanics and fluid dynamics. For example, they play a pivotal role in the thermal analysis of a system by formulating a general heat equation that can be analyzed using various appropriate boundary and initial conditions \cite{ref_book1}. Similarly, this kind of thermal analysis is a foremost step in the design of many safety-critical applications, such as aerospace, nuclear power plants and automobile engines.

The phenomenon of heat transfer/propagation can occur by three different means, namely, heat conduction \cite{ref_book1}, convection \cite{ref_book2}, and thermal radiation \cite{ref_book4f}. Heat conduction or diffusion is the flow of energy in a system/body from the region of high temperature to the region of low temperature by direct collision of molecules. Whereas, convection refers to the transfer of the energy due to the physical movement of a bulk fluid. Thermal radiation is the transfer of energy in the form of electromagnetic wave. Heat conduction is the most important type of heat transfer and it is commonly used to analyze problems arising in the design and operation of industrial appliances, such as heat exchanger and compressors. The first step for analyzing the heat conduction in a given system/body is to construct a mathematical model of the dynamics of the system, such as heat distribution  using the heat equation, which is a PDE. These dynamics provide the variation of the temperature as a function of position/space and time within the heat conducting system/body. The heat distribution (temperature field) usually depends on boundary conditions, initial conditions, material properties, and the geometry of the body. The next step in the heat conduction analysis is to find the solution of the heat equation modeled in the first step that can be obtained by determining a temperature distribution that is consistent with the initial and boundary conditions.

Heat equations are generally analyzed using numerical techniques \cite{ref_book5f} or analytical methods \cite{ref_book6f}. The two most widely used numerical techniques for analyzing PDE based heat equations are Finite Difference \cite{ref_finite_diff} and Finite Element \cite{ref_finite_elem} methods. These methods can solve the complex heat conduction problems by providing the closed-from solutions. However, they involve approximation and rounding of the associated mathematical expressions and thus cannot ensure absolute correctness of the results of the associated analysis. Unlike numerical solutions, the analytical methods for analyzing heat conduction do not involve any approximation of the associated mathematical expressions and thus are preferred on numerical methods for ensuring the correctness of the results. Some commonly used analytical techniques for solving the heat conduction problem are separation of variables \cite{ref_book7f} and transform methods \cite{ref_book8f}.

Conventionally, the heat conduction problem has been analyzed using paper-and-pencil proof and computer based numerical and symbolic methods. However, the former is human-error prone and it is not well-suited for large systems involving extensive human manipulation. Moreover, the required assumptions are not all explicitly mentioned in the analysis, which may lead to inaccurate results. Similarly, the numerical and symbolic methods are based on approximation of the mathematical results due to the finite precision of computer arithmetic. Moreover, the core of the tools involved in the symbolic methods based analysis has a large number of unverified algorithms that puts a question mark on the accuracy of the associated analysis. Given, the safety-critical nature of many systems, these conventional techniques cannot ensure absolute accuracy of the analysis.

As an alternative to related methods and tools, in this paper, we propose to use higher-order-logic theorem proving \cite{ref_book4} for formally analyzing the heat conduction problem in rectangular coordinates and thus overcome the above-mentioned inaccuracy limitations. In particular, we formally model the heat equation using the multivariable calculus theories of the HOL Light theorem prover capturing the heat conduction in the system/body. Next, to formally analyze the heat equation, we use the separation of variables method \cite{ref_book7f} to formally verify the solution of the PDE by incorporating all relevant boundary and initial conditions. One of the primary reasons for choosing HOL Light for the proposed work is the availability of rich theories of multivariable calculus, such as differential, integration, transcendental and real analysis. The HOL Light codes of our formalization is available at \cite{h_light}.

The remainder of the paper is structured as follows: In Section \ref{section2}, we provide an overview of related work on differential equations based formal analysis. Section \ref{section3} describes some fundamentals of the multivariate analysis libraries of the HOL Light theorem prover that are necessary for understanding the rest of the paper. We provide the formalization of the heat equation in rectangular coordinates in Section \ref{section4}. Section \ref{section5} presents the formal verification of the solution of the heat equation. Finally, Section \ref{section6} concludes the paper.

\section{Related Work}
\label{section2}
Many higher-order-logic theorem provers, such as HOL Light\footnote{\url{https://www.cl.cam.ac.uk/ jrh13/hol-light/}}, HOL4\footnote{\url{https://hol-theorem-prover.org/}}, Isabelle/\\HOL\footnote{\url{https://isabelle.in.tum.de/}}, Coq\footnote{\url{https://coq.inria.fr/}} and Mizar\footnote{\url{ http://www.mizar.org/}} have been used for the differential equations based formal analysis of the engineering and physical systems. For instance, Immler et al. \cite{ref_lncs1} used Isabelle/HOL for formally verifying the numerical solutions of Ordinary Differential Equation (ODE). The authors formalized the Initial Value Problems (IVPs) and formally verified the existence of a unique solution of the ODE. Moreover, the authors provide an approximation of the solution using the Euler's method. Immler et al. \cite{ref_article1} presented a formal reasoning support about the flow of ODEs using Isabelle/HOL. In particular, the authors formally verified a solution of ODEs incorporating various initial conditions. They also formalized the Poincaré map and formally verified its differentiability. However, both these approaches rely on approximating the solutions of differential equations representing the dynamical behavior of the underlying system. Guan et al. \cite{ref_article2} used the HOL Light theorem prover to formalize the Euler-Lagrange equation set that is based on Gâutex derivatives. In addition, the authors used their proposed formalization for formally verifying the least resistance problem of gas flow. Similarly, Sanwal et al. \cite{sanwal} formally verified the solutions of the second-order homogeneous linear differential equations using the HOL4 theorem prover. Moreover, they used their proposed formalization for formally verifying the damped harmonic oscillator and a second-order op-amp circuit. Rashid et al. formalized the Laplace \cite{rashid1} and the Fourier \cite{rashid2} transforms using HOL Light and used these formalization for differential equations based analysis of many systems, such as automobile suspension system \cite{rashid2}, unmanned free-swimming submersible vehicle \cite{rashid3} and platoon of automated vehicles \cite{rashid4}. However, the existing formalization of ODEs in HOL4 and HOL Light, respectively, do not provide the formalization of the solution when dealing with separable linear partial differential equations.

Boldo et al. \cite{ref_bol} utilized the Coq theorem prover for formally verifying the numerical solution of one-dimensional acoustic wave equation. The authors used the second-order centered finite difference scheme, commonly known as the three-point scheme for convergence of the result. Similarly, Boldo et al. \cite{ref_article_boldo} mechanically verified the correctness of a C program implementing numerical scheme for the solution of PDE using both automated and interactive theorem provers. Despite important contributions, both these works approximate the solution of acoustic wave equation and did not provide analytical solution.  Otsuki et al. \cite{ref_article4} formalized the method of separation of variables and superposition principle and used it for analyzing a one-dimensional wave equation using the Mizar theorem prover. However, they did not extend the solution for the infinite series. In the work we propose in this paper, we provide, for the first time, the formalization in HOL of the heat equation, in the form of a PDE modeling temperature variation for a rectangular solid. We conduct the formal verification in HOL Light of useful properties of the heat equation as well as verify its infinite series solution.

\section{Preliminaries}
\label{section3}
In this section, we provide an overview of some of the fundamental formal definitions and notations of the multivariate calculus theories of HOL Light that are necessary for understanding the rest of the paper. The derivative of a real-valued function is defined in HOL Light as follows:

\begin{definition}
\label{DEF:real_deriv} \emph{\textit{Real Derivative}} \\
{\small\textup{\texttt{$\vdash$ $\forall$f x.
    real\_derivative f x =  (@f'. (f has real derivative f') (atreal x))}}}
  \end{definition}
The function \texttt{real\_derivative} accepts a real valued function \texttt{f} that needs to be differentiated and a real number \texttt{x}, and provides the derivative of \texttt{f} with respect to \texttt{x}. It is formally represented in functional form using the Hilbert choice operator @. The function \texttt{has\_real\_derivative} expresses the same functionality in relational style.

\begin{definition}
\label{DEF:higher_real_deriv} \emph{\textit{Higher Real Derivative}} \\
{\small\textup{\texttt{$\vdash$ $\forall$f x.
     higher\_real\_derivative 0 (f:real$\rightarrow$real) (x:real) =  f x $\wedge$ \\
  \hspace*{0.4cm}   (!n. higher\_real\_derivative (SUC n) (f:real$\rightarrow$real) (x:real) = \\
\hspace*{2.2cm} (real\_derivative ($\lambda$x. higher\_real\_derivative n f x) x))}}}
  \end{definition}
The HOL Light function \texttt{higher\_real\_derivative} accepts an order \texttt{n} of the derivative, a real-valued function \texttt{f} and a real number \texttt{x}, and provides a higher-order derivative of order \texttt{n} for the function \texttt{f} with respect to \texttt{x}.

The infinite summation over a function \texttt{f}: $\mathbb{N}$ $\rightarrow$ $\mathbb{R}$ is formalized in HOL Light as follows:

\begin{definition}
\label{DEF:real_sums} \emph{\textit{Real Sums}} \\
{\small\textup{\texttt{$\vdash$ $\forall$s f L.
    real\_sums (f real\_sums l) s $\Leftrightarrow$ \\
   \hspace*{1.2cm}   (($\lambda$n. sum (s INTER (0..n)) f) $\rightarrow$ l) sequentially}}}
  \end{definition}
The HOL Light function \texttt{real\_sums} accepts a set of natural numbers \texttt{s}: \texttt{$\mathbb{N}$} $\rightarrow$ \texttt{bool}, a function \texttt{f}: $\mathbb{N}$ $\rightarrow$ $\mathbb{R}$ and a limit value \texttt{l}: $\mathbb{R}$, and returns the traditional mathematical expression $\sum\limits_{k = 0}^{\infty}{f(k)} = L$. Here, \textsf{\texttt{INTER}} captures the intersection of two sets. Similarly, \texttt{sequentially} represents a net providing a sequential growth of a function $f$, i.e., $f(k), f(k + 1), f(k + 2), . . . ,$ etc. This is mainly used in modeling the concept of an infinite summation.

We provide the formalization of the summability of a function \texttt{f}: $\mathbb{N}$ $\rightarrow$ $\mathbb{R}$ over \texttt{s}: $\mathbb{N}$ $\rightarrow$ \texttt{bool}, which ensures that there exist some limit value \texttt{L}: $\mathbb{R}$, such that $\sum\limits_{k = 0}^{\infty}{f(k)} = L$ in HOL Light as:

\begin{definition}
\label{DEF:real_summability} \emph{\textit{Real Summability}} \\
{\small\textup{\texttt{$\vdash$ $\forall$s f.
    real\_summable s f = ?l. (f real\_sums l)}}}
  \end{definition}

Now, we provide a formalization of an infinite summation, which will be used in the formal analysis of the heat conduction problem in Section \ref{section5} of the paper.

 \begin{definition}
\label{DEF:real_infsum} \emph{\textit{Real Infsum}} \\
{\small\textup{\texttt{$\vdash$ $\forall$s f.
    real\_infsum s f = @l. (f real\_sums l) s}}}
    \label{dp5}
  \end{definition}

\noindent where the HOL Light function \texttt{real\_infsum} accepts \texttt{\textsf{s}}: \texttt{\textsf{num}} $\rightarrow$ \texttt{\textsf{bool}} specifying the starting point and a function \texttt{f} of data-type $\mathbb{N}$ $\rightarrow$ $\mathbb{R}$, and returns a limit value \texttt{l}: $\mathbb{R}$  to which the infinite summation of \texttt{f} converges from the given \texttt{s}.

An infinite summation of a real-valued function Definition \ref{dp5} can be mathematically expressed in an alternate form as follows:

\begin{equation*}
 \sum_{w=0}^{\infty}f_{w}(x) = \lim_{N \to \infty}\sum_{w=0}^{N}f_{w}(x)
\end{equation*}
We proved this equivalence in HOL Light as follows:

\begin{theorem}
\label{THM:alternative_infsum} \emph{\textit {Alternate Representation of an Infinite Summation}} \\
\textup{{\small
\texttt{$\vdash$ $\forall$f k s.  real\_infsum s ($\lambda$w. f w x) = \\
\hspace*{1.5cm} reallim sequentially ($\lambda$k. sum (s INTER (0..k))($\lambda$w. f w x))}}}
\end{theorem}

\section{Formalization of the Heat Conduction Problem}
\label{section4}
Heat conduction is a phenomenon of energy transfer that occurs due to differences in temperature in adjacent components of a body/system. The heat is transferred from the high-temperature side to the low-temperature side  until the body reaches its thermal equilibrium. The heat conduction or temperature variation can be mathematically defined as a function of space and time. Generally, the heat conduction in a body is three dimensional  i.e., the conduction is significant in all three dimensions and a temperature variation in a body can be modeled as $T = T (x, y, z, t)$. The heat conduction is said to be two-dimensional when the conduction is significant in two-dimensions and negligible in the third dimension. Similarly, it is one-dimensional when the conduction is significant in one-dimensional only and the temperature variable can be modeled as $T = T (x, t)$. In this paper, we focus on the formalization of the one-dimensional heat conduction problem. In particular, we formally model the temperature variation in a rectangular slab using a PDE as a heat equation and formally verify its analytical solution by the method of separation of variables based on various boundary and initial conditions.

\subsection{Heat Conduction Problem Formulation}
\begin{figure}[ht!]
    \centering
    \includegraphics[width=0.4 \textwidth]{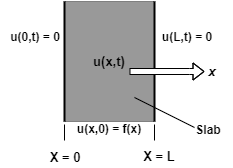}
\caption{Heat Conduction Across Thickness of a Slab \cite{ref_book_app}}
    \label{figure1}
    \end{figure}

A heat conduction problem for a rectangular slab having a thickness $L$ is depicted in Figure \ref{figure1}. We consider it as a one-dimensional heat conduction problem. Here, the function $u (x, t)$ provides the temperature in the slab at a point $x$ and time $t$ \cite{ref_book4s}.

We can mathematically express the one-dimensional heat conduction (temperature variation) in the rectangular slab as follows \cite{ref_book_app}:
\begin{equation}
\frac{\partial u(x,t)}{\partial t} = c \dfrac{\partial^{2}u(x,t)}{\partial x^{2}} \quad\quad 0 < x < L, \quad t > 0
\label{1}
\end{equation}
where c is the thermal diffusivity of the slab that depends on the material used for constructing the slab. Equation (\ref{1}) can be equivalently written as:
\begin{equation*}
    \frac{\partial u(x,t)}{\partial t} - c \dfrac{\partial^{2}u(x,t)}{\partial x^{2}} = 0
\end{equation*}

Moreover, the solution of the heat equation (Equation (\ref{1})) should satisfy the following initial and boundary conditions.\\

Initial Condition:
\begin{equation}
u (x,t) \mid_{t = 0} = u (x,0) = f(x)
\label{2}
\end{equation}
\hspace*{0.35cm} Boundary Conditions:
\begin{equation}
u (x,t) \mid_{x = 0} = u (0,t) = 0
\label{3}
\end{equation}
\begin{equation}
u (x,t) \mid_{x = L} = u (L,t) = 0
\label{4}
\end{equation}
The heat equation (Equation (\ref{1})) along with Equations (\ref{2}), (\ref{3}) and (\ref{4}) is known as the initial boundary-value problem. It becomes an initial-value problem with respect to time that considers the only initial condition represented by Equation (\ref{2}). Whereas, in the case of its dependence on space only, it represents a boundary-value problem by incorporating the two boundary conditions expressed as Equations (\ref{3}) and (\ref{4}).  Next, to formally verify the solution of the heat equation, we need to formalize it in higher-order logic.

\subsection{Formalization of the Heat Equation}
We formalize the heat equation (Equation (\ref{1})) capturing the one-dimensional heat conduction in a rectangular slab in HOL Light as follows:
\begin{definition}\emph{\textit{The Heat equation}}{\small \\
\textup{\texttt{$\vdash$ heat\_equation u(x,t) c $\Leftrightarrow$ heat\_operator u (x,t) c = \&0}}}
\label{d6}
  \end{definition}
where \texttt{heat\_equation} accepts a function \texttt{u} of type ($\mathbb{R} \times \mathbb{R} \rightarrow \mathbb{R})$, a space variable \texttt{x: $\mathbb{R}$}, a time variable \texttt{t: $\mathbb{R}$} and the thermal diffusivity constant \texttt{c}, and returns the corresponding heat equation. The function \texttt{heat\_operator} is formalized as follows:

\begin{definition}\emph{\textit{Heat operator}}{\small \\
\textup{\texttt{$\vdash$ $\forall$u x t. \\
heat$_{-}$operator u (x,t) c = higher\_real\_derivative 1  ($\lambda$t. u (x,t)) t - \\
\hspace*{3.6cm} c * higher\_real\_derivative 2 ($\lambda$x. u (x,t)) x}}}
\label{d7}
 \end{definition}
Next, we verify a few important properties of the \texttt{heat\_operator} Definition \ref{d7} that are required in formally verifying the solution of the heat equation.
\\
\\
\begin{theorem}
\label{THM:linearity} \emph{\textit{Linearity}} \\
\textup{{\small
\texttt{$\vdash$ $\forall$u x t a b.
\begin{enumerate}
\setlength\itemindent{0.23cm}
\item [{[A1]}] ($\forall$t. ($\lambda$t. u (x,t)) real\_differentiable atreal t) $\wedge$
\item [{[A2]}] ($\forall$t. ($\lambda$t. v (x,t)) real\_differentiable atreal t) $\wedge$
\item [{[A3]}] ($\forall$x. ($\lambda$x. u (x,t)) real\_differentiable atreal x) $\wedge$
\item [{[A4]}] ($\forall$x. ($\lambda$x. v (x,t)) real\_differentiable atreal x) $\wedge$
\item [{[A5]}] ($\forall$x. ($\lambda$x. real\_derivative ($\lambda$x. u (x,t)) x) \\
\hspace*{1cm} real\_differentiable atreal x) $\wedge$
\item [{[A6]}] ($\forall$x. ($\lambda$x. real\_derivative ($\lambda$x. v (x,t)) x)  \\
\hspace*{1cm}  real\_differentiable atreal x)
\end{enumerate}
$\Rightarrow$ (heat\_operator ($\lambda$(x,t). u (x,t) + v (x,t)) (x,t) c =\\
\hspace*{1.0cm}heat\_operator ($\lambda$(x,t). u (x,t)) (x,t) c +\\ \hspace*{1.2cm} heat\_operator ($\lambda$(x,t). v (x,t)) (x,t) c)}}}
\end{theorem}
Assumptions \texttt{A1} and \texttt{A2} ensure that the real-valued functions \texttt{u} and \texttt{v} are differentiable at \texttt{t}, respectively. Assumptions \texttt{A3} and \texttt{A4} assert the differentiability of the functions \texttt{u} and \texttt{v} at \texttt{x}, respectively. Similarly, Assumptions \texttt{A5} and \texttt{A6} provide the differentiability conditions for the derivatives of the functions \texttt{u} and \texttt{v} at \texttt{x}, respectively. The proof of the above theorem is mainly based on the properties of derivative and differentiability of real-valued functions.

\begin{theorem}
\label{THM:scalar} \emph{\textit{Scalar Multiplication}} \\
\textup{{\small
\texttt{$\vdash$ $\forall$u x t a.
\begin{enumerate}
\setlength\itemindent{0.23cm}
\item [{[A1]}] ($\forall$t. ($\lambda$t. u (x,t)) real\_differentiable atreal t) $\wedge$
\item [{[A2]}] ($\forall$x. ($\lambda$x. u (x,t)) real\_differentiable atreal x) $\wedge$
\item [{[A3]}] ($\forall$x. ($\lambda$x. real\_derivative ($\lambda$x. u (x,t)) x)\\
\hspace*{1.2cm}real\_differentiable atreal x)
\end{enumerate}
$\Rightarrow$ heat\_operator ($\lambda$(x,t). a * u (x,t)) (x,t) c = \\
\hspace*{3.5cm} a * heat\_operator ($\lambda$(x,t). u (x,t))(x,t) c}}}
\end{theorem}
Assumptions \texttt{A1} and \texttt{A2} ensure that the real-valued function \texttt{u} is differentiable at \texttt{t} and \texttt{x}, respectively. Assumption \texttt{A3} asserts the differentiability condition for the derivative of the function \texttt{u}.

\section{Formal Verification of the Solution of the Heat Equation}
\label{section5}
To find out the solution of the boundary-value problem, i.e., heat equation alongside the boundary conditions Equations (\ref{1}), (\ref{3}) and (\ref{4}), we use the method of separation of variables that reduces the problem of solving a partial differential equation to a problem of solving the equivalent ordinary differential equations. By this method, we can mathematically express the solution of the heat equation $u (x, t)$ as a separable equation as follows:
\begin{equation}
    u(x,t) = X(x)W(t)
    \label{5}
\end{equation}
\noindent where $X$ and $W$ are functions of $x$ and $t$, respectively. We formalize Equation (\ref{5})  in HOL Light as follows:

\begin{definition}\emph{\textit{Separable}}{\small \\
    \textup{\texttt{$\vdash$ $\forall$u X W t x. separable u x t X W =  X(x) * W(t)}}}
  \end{definition}
By using Equation (\ref{5}) in the heat equation (Equation (\ref{1})) and after simplification, we obtain the following equation.
\begin{equation}
\dfrac{1}{c}\dfrac{\partial [X(x)W(t)]}{\partial t} = \dfrac{\partial ^{2} [X(x)W(t)]}{\partial x^{2}}
\label{6}
\end{equation}
Next, using the property of the partial derivative of a separable function transforms the above equation as follows:
\begin{equation}
\dfrac{1}{c}\dfrac{dW(t)}{dt}X(x) = W(t)\dfrac{d^{2}X(x)}{dx^{2}}
\label{7}
\end{equation}
\noindent where the operator $\frac{d}{dt}$ captures the simple derivative with respect to $t$. We formally verify the equivalence of the left-hand-sides of Equations (\ref{6}) and (\ref{7}) as the following HOL Light theorem.
\begin{theorem}
\label{THM:equivalence} \emph{\textit{Equivalence of Partial and Simple Derivatives (Left-hand Side)}}\\
\textup{{\small
\texttt{$\vdash$ $\forall$u X x W t.
\begin{enumerate}
\setlength\itemindent{0.23cm}
\item [{[A1]}] (X real\_differentiable atreal t) $\wedge$
\item [{[A2]}] (W real\_differentiable atreal t)
\end{enumerate}
\hspace*{0.2cm} $\Rightarrow$
(real\_derivative ($\lambda$t. separable u x t X W) t) = \\ \hspace*{0.8cm} real\_derivative W t * X x
}}}
\label{t4}
\end{theorem}
Assumptions \texttt{A1} and \texttt{A2} provide the differentiability of the functions \texttt{X} and \texttt{W} at \texttt{t}, respectively. The proof process of the above theorem is mainly based on the properties of derivatives and differentiability of the real-valued functions alongwith some arithmetic reasoning. Similarly,  we formally verify the equivalence of the right-hand-sides of Equations (\ref{6}) and (\ref{7}) as follows:
\begin{theorem}
\label{THM:eqv_right} \emph{\textit{Equivalence of Partial and Simple Derivatives (Right-hand Side)}} \\
\textup{{\small
\texttt{$\vdash$ $\forall$u X x W t.
\begin{enumerate}
\setlength\itemindent{0.23cm}
\item [{[A1]}] ($\forall$x. X real\_differentiable atreal x) $\wedge$
\item [{[A2]}] ($\forall$x. W real\_differentiable atreal x) $\wedge$
\item [{[A3]}] ($\lambda$x. real$_{-}$derivative X x) real\_differentiable atreal x
\end{enumerate}
$\Rightarrow$ higher\_real\_derivative 2  ($\lambda$x. (separable u x t X W)) x = \\
\hspace*{3.5cm} W t * higher\_real\_derivative 2 ($\lambda$x. X x) x
}}}
\label{t5}
\end{theorem}
Assumptions \texttt{A1} and \texttt{A2} are very similar to that of Theorem \ref{t4}. Assumption \texttt{A3} ensures that the first-order derivative of the real-valued function \texttt{X} is differentiable at \texttt{x}. The verification of Theorem \ref{t5} is similar to that of Theorem \ref{t4}.

Now, after rearranging various terms, Equation (\ref{7}) can be expressed as follows:
\begin{equation}
\dfrac{1}{c}\dfrac{dW(t)}{dt}\dfrac{1}{W(t)} =  \dfrac{1}{X(x)}\dfrac{d^{2}X(x)}{dx^{2}} = -\beta^{2}
\label{8}
\end{equation}
\noindent where the left- and right-hand sides are functions of only $t$ and $x$, respectively. The equivalence of these two functions of different variables is only possible when both are equal to some constant, which is represented by $- \beta^2$ in the above equation.

The above equation can be equivalently represented by the following two ordinary differential equations.
\begin{equation}
    \dfrac{d^{2}X(x)}{dx^{2}} + \beta^{2}X(x) = 0
    \label{9}
\end{equation}
and
\begin{equation}
    \dfrac{dW(t)}{dt} + c. \beta^{2}W(t) = 0
    \label{10}
\end{equation}
Now, our problem of solving a boundary-value problem Equations (\ref{1}), (\ref{3}) and (\ref{4}) has been transformed to solving a set of linear homogenous differential equations with constant coefficients Equations (\ref{9}) and (\ref{10}). Moroever, the solution of the heat equation Equation (\ref{1}) can be obtained by multiplying the solution of these two equations. \\
The solution of Equation (\ref{9}) is mathematically expressed as:
\begin{equation}
   X(x) = Acos(\beta x) + Bsin(\beta x)
   \label{11}
\end{equation}
\noindent where $A$ and $B$  are the arbitrary constants that can be computed by applying the boundary conditions. Similarly, the solution of the second differential equation Equation (\ref{10}) is mathematically described as:
\begin{equation}
   W(t) = Ce^{-\beta^{2}ct}
   \label{12}
\end{equation}
where $C$ is the constant of integration and can be computed by applying the boundary conditions. \\
We formalize the two differential equations Equations (\ref{9}) and (\ref{10}) in HOL Light as follows:
\begin{definition}\emph{\textit{Formalization of Equation (9)}}{\small \\
    \textup{\texttt{$\vdash$ $\forall$X x b.
    first\_equation X x b $\Leftrightarrow$ \\
    \hspace*{0.3cm} higher\_real\_derivative 2 ($\lambda$x.  X (x)) x +
b pow (2) * ($\lambda$x. X(x)) x = 0}}}
\label{d9}
  \end{definition}
\begin{definition}\emph{\textit{Formalization of Equation (10)}}{\small \\
    \textup{\texttt{$\vdash$ $\forall$W t b c.\\
  \hspace*{0.3cm}  second\_equation W t b c $\Leftrightarrow$ \\
  \hspace*{1.5cm}  real\_derivative ($\lambda t.$ W (t)) t + c * b pow (2) * W(t) = 0}}}
    \label{d10}
  \end{definition}
Similarly, we formalized the solutions of these differential equations in HOL Light as:
\begin{definition}\emph{\textit{Solution of First Differential Equation} }{\small \\
    \textup{\texttt{$\vdash$ $\forall$A B x b.
 first\_equation\_sol A B x b =  A * cos(b * x) + B * sin(b * x)}}}
 \label{d11}
  \end{definition}

\begin{definition}\emph{\textit{Solution of Second Differential Equation} }{\small \\
    \textup{\texttt{$\vdash$ $\forall$C c b t.\\
  \hspace*{0.3cm} second\_equation\_sol C c b t = C * exp (--c * b pow (2) * t)}}}
 \label{d12}
  \end{definition}
Next, we formally verify the solution of the first differential equation Equation (\ref{9}) as the following HOL Light theorem:

\begin{theorem}
\label{THM:sol_diff1} \emph{\textit{Solution of First Differential Equation}} \\
\textup{{\small
\texttt{$\vdash$ $\forall$A B x b.\\
 \hspace*{0.3cm} (first\_equation ($\lambda x.$ first\_equation\_sol A B x b)) x b}}}
\label{t6}
\end{theorem}
The proof process of the above theorem is based on Definitions \ref{d9} and \ref{d10} and properties of real derivative alongside some real arithmetic reasoning.\\
Similarly, we formally verify the solution of the second differential equation Equation (\ref{10}) as follows:
  \begin{theorem}
\label{THM:sol_diff2} \emph{\textit{Solution of Second Differential Equation}} \\
\textup{{\small
\texttt{$\vdash$ $\forall$C c b t.\\
 \hspace*{0.3cm} (second\_equation ($\lambda$t. second\_equation\_sol C c b t))(t) b c}}}
\label{t7}
\end{theorem}
The proof process of the above theorem is based on Definitions  \ref{d10} and \ref{12} and properties of real derivative alongside some real arithmetic reasoning.

To find out the values of arbitrary constants $A$ and $B$ of the solution of the ordinary differential equation expressed as Equation (\ref{11}), we apply the corresponding boundary conditions. Applying the first boundary condition Equation (\ref{3}) results into $A = 0$. Similarly, the application of the second boundary condition Equation \ref{4} provides $B sin (\beta L) = 0$. We formally verify values of these arbitrary constants based on the corresponding boundary conditions in HOL Light as follows:
\begin{theorem}
\label{THM:arb_a} \emph{\textit{Verification of the Arbitrary Constant A}} \\
\textup{{\small
\texttt{$\vdash$ $\forall$A B x b.\\
 \hspace*{0.8cm} x = \&0 $\wedge$ first\_equation\_sol A B x b = \&0 \\
\hspace*{6.0cm} $\Rightarrow$ A = \&0}}}
\label{t8}
\end{theorem}

\begin{theorem}
\label{THM:arb_b} \emph{\textit{Verification of the Arbitrary Constant B}} \\
\textup{{\small
\texttt{$\vdash$ $\forall$A B x b L.\\
 \hspace*{0.8cm} x = L $\wedge$ A = \&0 $\wedge$ first\_equation\_sol A B x b = \&0 \\
 \hspace*{0.9cm}  $\Rightarrow$ first\_equation\_sol x b A B = B * sin(b * L)}}}
\label{t9}
\end{theorem}
The equation $B sin (\beta L) = 0$ holds if $B = 0$ or $sin (\beta L) = 0$. In case of $B = 0$ alongside $A = 0$, it results into $X (x) = 0$. This further provides $u (x, t) = 0$ as a solution to the heat equation, which is an uninteresting trivial solution. This means that $B$ is equal to some non-zero value, which implies that $sin (\beta L) = 0$. Since $\beta$ can have infinitely many values for which $sin (\beta L) = 0$ holds, namely $\beta = \beta_w = \frac{\omega \pi}{L}$. This results into a non-trivial solution of the boundary-value problem as follows:

\begin{equation}
 u (x,t) = u_{w}(x,t) = \left[B_{w}sin\left(\dfrac{w\pi x}{L}\right)\right]e^{-\left(\dfrac{w\pi}{L}\right)^{2}ct}
 \label{13}
\end{equation}
Now, assume that the function $f(x)$ in initial condition Equation (\ref{2}) is a linear combination of the function $sin (\frac{w \pi x}{L})$, i.e., Fourier sine series representation as follows:
\begin{equation}
    f(x) = \sum_{w=1}^{\infty}B_{w}sin\left(\dfrac{w\pi x}{L}\right) \label{14}
\end{equation}
We can mathematically express the general solution of the heat equation as the following equation since it is a linear combination of the non-trivial solutions of the boundary-value problem that satisfies the initial condition expressed as Equation (\ref{14}).
\begin{equation}
u(x,t) = \sum_{w=1}^{\infty}u_{w} (x,t) = \sum_{w=1}^{\infty}B_{w}sin \left(\dfrac{w\pi x}{L}\right)e^{-\left(\dfrac{w\pi}{L}\right)^{2}ct}
\label{15}
\end{equation}
The constant $B_w$ of the Fourier sine series representation of $f(x)$ can be determined using the orthogonality property of the sine function and is mathematically expressed as follows:

\begin{equation}
    B_{w} = \dfrac{2}{L}\int_{0}^{L} f(x)sin \left(\dfrac{w\pi x}{L}\right)dx \quad\quad\quad w = 1,2,3...
    \label{16}
\end{equation}
We formalize the Fourier sine coefficient in HOL Light as follows:
\begin{definition}
\emph{\textit{Fourier Sine Coefficient}}{\small \\ \textup{\texttt{$\vdash$ $\forall$f w L.\\
 \hspace*{0.3cm} fourier\_sine\_coefficient f w L = \\
 \hspace*{0.9cm}  2 / L * (real\_integral (real\_interval [0,L])($\lambda$x. (f x) * \\
 \hspace*{1.5cm} sin (\&w * pi * x / L)))}}}
\label{d13}
  \end{definition}
where \texttt{fourier\_sine\_coefficient} accepts a function \texttt{f : $\mathbb{R}$ $\rightarrow$ $\mathbb{R}$}, a number \texttt{w} and the width of the slab \texttt{L}, and returns a real number representing the Fourier sine coefficient of the function \texttt{f}.\\

Now, the solution of the heat equation capturing the heat conduction in a rectangular slab can be alternatively expressed as:
\begin{equation}
u (x,t) = \sum_{w=1}^{\infty}u_{w} (x,t) = \sum_{w=1}^{\infty}\left(\dfrac{2}{L}\int_{0}^{L} f(x)sin \left(\dfrac{w\pi x}{L}\right)dx \right) sin \left(\dfrac{w\pi x}{L}\right)e^{-\left(\dfrac{w\pi}{L}\right)^{2}ct}
\label{17}
\end{equation}

We formalize the generalized solution of the heat equation (Equation (\ref{17})) in HOL Light as follows:
\begin{definition}
\emph{\textit{Generalized Solution of the Heat Equation}}{\small \\
    \textup{\texttt{$\vdash$ $\forall$f x t c L.\\
 \hspace*{0.3cm}  heat\_solution f x t c L = real\_infsum (from 1) \\
 \hspace*{0.9cm} ($\lambda $w. (fourier\_sine\_coefficient f w L) * \\
 \hspace*{1.2cm} exp (--c * ((\&w * pi / L) pow 2) * t) *  sin (\&w * pi * x / L))}}}
 \label{d14}
  \end{definition}

The convergence of the generalized solution of the heat equation depends on the convergence of the infinite series $u_w (x, t)$ and is mathematically expressed as the following bound on $u_w (x, t)$.

\begin{equation}
   \vert u_{w} (x,t) \vert \leq M_{w}
\end{equation}
where
\begin{equation}
M_{w} = \left(\dfrac{2}{L}\int_{0}^{L} \vert f(x) \vert dx \right)e^{-\left(\dfrac{w\pi}{L}\right)^{2}ct}
\end{equation}
We compute the upper bound $M_w$ using the upper bound on the Fourier coefficient $B_w$, and the fact that $\left \vert sin\left(\dfrac{w \pi x}{L}\right) \right \vert \leq 1$, along with the following property of the integral:
\begin{equation}
    \left \vert \int_{a}^{b} f(x)dx  \right \vert \leq \int_{a}^{b} \vert f(x)\vert dx.
\end{equation}
Next, we formally verify the convergence of the generalized solution of the heat equation as the following HOL Light theorem.

\begin{theorem}
\label{THM:conv_gen} \emph{\textit{Convergence of the Generalized Solution}} \\
\textup{{\small
\texttt{$\vdash$ $\forall$f x c L t.
\begin{enumerate}
\setlength\itemindent{0.23cm}
\item [{[A1]}] \&0 < L $\wedge$ [A2] \&0 < t $\wedge$ [A3] \&0 < c  $\wedge$
\item [{[A4]}] f absolutely\_real\_integrable\_on real\_interval [\&0, L]
\end{enumerate}
$\Rightarrow$ (($\lambda$w. fourier\_sine\_coefficient f w L * \\
\hspace*{1.7cm}exp (--c * (\&w * pi / L) pow 2 * t)  sin (\&w * pi * x / L)) \\
\hspace*{3.7cm}real\_sums heat\_solution f x t c L) (from 1)}}}
\label{t10}
\end{theorem}
Assumptions (\texttt{A1-A3}) ensure that  the length \texttt{L}, the time \texttt{t} and the constant \texttt{c} are positive real values. Assumption (\texttt{A4}) provides the absolute integrability of the function \texttt{f} over the interval \texttt{[0,L]}. The conclusion presents the convergence of the generalized solution of the heat equation. The verification of above Theorem \ref{t10} is mainly based on the following two important lemmas about the summability of the bound $M_w$ and the generalized solution alongside some real arithmetic reasoning.
\begin{lemma}
\label{THM:sum_bound} \emph{\textit{Summability of the Bound $M_{w}$}} \\
\textup{{\small
\texttt{$\vdash$ $\forall$f c L t.
\begin{enumerate}
\setlength\itemindent{0.23cm}
\item [{[A1]}] \&0 < L $\wedge$ [A2] \&0 < t $\wedge$
[A3] \&0 < c
\end{enumerate}
$\Rightarrow$ real\_summable (from 1) ($\lambda$w. \&2 / L * real\_integral \\
\hspace*{2.9cm}(real\_interval [\&0,L]) ($\lambda$x. abs (f x)) * \\
\hspace*{4.8cm} exp (--c * ((\&w * pi / L) pow 2) * t))}}}
\label{l1}
\end{lemma}
Assumptions (\texttt{A1-A3}) are the same as those of Theorem \ref{t10}. The conclusion of the above lemma provides the summability of the upper bound $M_w$. The verification of Lemma \ref{l1} is mainly based on the Ratio test \cite{calculus} along with some real arithmetic reasoning.

\begin{lemma}
\label{THM:sum_gen} \emph{\textit{Summability of the Generalized Solution}} \\
\textup{{\small
\texttt{$\vdash$ $\forall$f x c L t.
\begin{enumerate}
\setlength\itemindent{0.23cm}
\item [{[A1]}] \&0 < L $\wedge$ [A2] \&0 < t $\wedge$ [A3] \&0 < c  $\wedge$
\item [{[A4]}] f absolutely\_real\_integrable\_on real\_interval [\&0, L]
\end{enumerate}
$\Rightarrow$ real\_summable (from 1)($\lambda$w. fourier\_sine\_coefficient f w L *  \\
\hspace*{1.3cm}exp (--c * (\&w * pi / L) pow 2 * t) * sin (\&w * pi * x / L))}}}
\label{l2}
\end{lemma}
Assumptions (\texttt{A1-A4}) are the same as those of Theorem \ref{t10}. The verification of Lemma \ref{l2} is mainly based on the Comparison test \cite{calculus} and Lemma \ref{l1} along with some real arithmetic reasoning. More details about the verification of these lemmas and the convergence of the generalized solution of the heat equation can be found in our HOL Light script \cite{h_light}.

Next, we formally verify some interesting properties involving the derivatives of the general solution with respect to position $x$ and time $t$ that capture the heat conduction (variation of temperature) in the rectangular slab with respect to position and time.
\begin{theorem}
\label{THM:deriv_gen} \emph{\textit{Derivative of the Generalized Solution with Respect to Time}} \\
\textup{{\small
\texttt{$\vdash$ $\forall$f x t c L u u'.
\begin{enumerate}
\setlength\itemindent{0.23cm}
\item [{[A1]}] ($\forall$t. (($\lambda$w. (fourier\_sine\_coefficient f w L) * \\
\hspace*{0.3cm}exp (--c * (\&w * pi / L) pow 2 * t) * sin (\&w * pi * x / L)) \\
\hspace*{0.3cm}real\_sums u(x,t)) (from 1)) $\wedge$
\item [{[A2]}] ($\forall$t. (($\lambda$w. --c * (\&w * pi / L) pow 2 * (fourier\_sine\_coefficient \\
\hspace*{0.2cm} f w L) * exp (--c * (\&w * pi / L) pow 2 * t) * \\
\hspace*{0.2cm} sin (\&w * pi * x / L)) real\_sums u'(x,t)) (from 1)) $\wedge$
\item [{[A3]}] (($\lambda$t. u(x,t)) has\_real\_derivative u'(x,t)) (atreal t)
\end{enumerate}
$\Rightarrow$ real\_derivative ($\lambda$t. heat\_solution f x t c L ) t =  \\
\hspace*{0.6cm} real\_infsum (from 1) ($\lambda$w. --c * (\&w * pi / L) pow 2) * \\
\hspace*{0.9cm}(fourier\_sine\_coefficient f w L) * \\
\hspace*{1.2cm} exp (--c * (\&w * pi / L) pow 2) * t) * sin (\&w * pi * x / L))}}}
\label{t11}
\end{theorem}
\vspace{-0.2cm}
Assumption \texttt{A1} provides the condition that the infinite series converges to the function $u (x, t)$. Similarly, Assumption \texttt{A2} asserts that the derivative of the infinite series with respect to $t$ converges to the derivative of function $u (x, t)$, i.e. $u'(x, t)$. Assumption \texttt{A3} ensures the function $u$ has derivative $u'(x, t)$ at point $t$. The verification of the above theorem is mainly based on swapping the operation of differentiation and infinite summation alongwith properties of the infinite summation and derivatives.

\begin{theorem}
\label{THM:first_deriv} \emph{\textit{First Derivative of the Generalized Solution with Respect to Space}} \\
\textup{{\small
\texttt{$\vdash$ $\forall$f x t c L u u'.
\begin{enumerate}
\setlength\itemindent{0.23cm}
\item [{[A1]}] ($\forall$x. (($\lambda$w. (fourier\_sine\_coefficient f w L) * exp (--c * (\&w * pi / L) \\
\hspace*{0.3cm}pow 2 * t) * sin (w * pi * x / L)) real\_sums u(x,t)) (from 1)) $\wedge$
\item [{[A2]}] (($\forall$x.(($\lambda$w. (fourier\_sine\_coefficient f w L) * exp (--c * (\&w * pi / L) \\
\hspace*{0.1cm} pow 2 * t) * (\&w * pi/ L) * cos (\&w * pi * x / L)) real\_sums \\
\hspace*{0.1cm} u'(x,t)))(from 1) $\wedge$
\item [{[A3]}] (($\lambda$x. u(x,t)) has\_real\_derivative u'(x,t)) (atreal x)
\end{enumerate}
$\Rightarrow$ real\_derivative ($\lambda$x. heat\_solution f x t c L) x = \\
\hspace*{0.5cm} real\_infsum (from 1)($\lambda$w. (fourier\_sine\_coefficient f w L) * \\
\hspace*{1.55cm}exp (--c * ((\&w * pi / L) pow 2) * t) * (\&w * pi / L) * \\
\hspace*{6.55cm}cos (\&w * pi * x / L)) }}}
\label{t12}
\end{theorem}
The proof process of Theorem \ref{t12} is very similar to that of Theorem \ref{t11}.
\begin{theorem}
\label{THM:second_der} \emph{\textit{Second Derivative of the General Solution with Respect to Space}} \\
\textup{{\small
\texttt{$\vdash$ $\forall$f x t c L u u' u''.
\begin{enumerate}
\setlength\itemindent{0.23cm}
\item [{[A1]}] ($\forall$x. (($\lambda$w. (fourier\_sine\_coefficient f w L) * exp (--c * (\&w * pi / L)\\
\hspace*{0.1cm} pow 2 * t) * sin (\&w * pi * x / L)) real\_sums u (x,t)) (from 1)) $\wedge$
\item [{[A2]}] (($\forall$x.(($\lambda$w. (fourier\_sine\_coefficient f w L) * exp (--c * (\&w * pi / L) \\
\hspace*{0.1cm} pow 2 * t) * (\&w * pi/ L) * cos (\&w * pi * x / L))  real\_sums \\
\hspace*{0.1cm} u'(x,t)))(from 1) $\wedge$
\item [{[A3]}] (($\lambda$x. u(x,t)) has$_{-}$real$_{-}$derivative u' (x,t)) (atreal x) $\wedge$
\item [{[A4]}] (($\forall$x. (($\lambda$w. (fourier\_sine\_coefficient f w L) * exp (--c * (\&w * pi / L) \\
\hspace*{0.1cm} pow 2 * t) * (\&w * pi / L) pow 2 *
--sin (\&w * pi * x / L)) real\_sums \\
\hspace*{0.1cm} u''(x,t))) (from 1) $\wedge$
\item [{[A5]}] (($\lambda$x. u'(x,t)) has\_real\_derivative
u''(x,t)) (atreal x)
\end{enumerate}
$\Rightarrow$ higher\_real\_derivative 2 ($\lambda$x. heat\_solution f x t c L) x = \\
\hspace*{0.5cm} real$_{-}$infsum (from 1)
($\lambda$w. (fourier\_sine\_coefficient f w L) * \\
\hspace*{0.7cm} exp (--c * ((\&w * pi / L) pow 2) * t) * ((\&w * pi / L) pow 2) *\\
\hspace*{0.9cm} --sin (\&w * pi * x / L))}}}
\label{t13}
\end{theorem}
\vspace{-0.1cm}
The verification of the above theorem is mainly based on Theorem \ref{t12} and properties of derivatives along with some arithmetic reasoning.
\vspace{0.1cm}

\subsection* {Discussion}
The distinguishing feature of our proposed formal analysis of the heat conduction problem, as compared to traditional analysis techniques, is that all verified theorems are of generic nature, i.e., all functions and variables involved in these theorems are universally quantified and thus can be specialized based on the requirement of the analysis of a rectangular slab with any width and corresponding boundary and initial conditions. Another advantage of our proposed approach is the inherent soundness of the theorem proving technique. It ensures that all the required assumptions are explicitly present along with the theorem, which are often ignored in conventional simulation based analysis and their absence may affect the accuracy of the corresponding analysis. One of the major difficulties in the proposed formalization was the swapping of the infinite summation and the differential operator that is used in the verification of Theorems \ref{t11}-\ref{t13}. The mathematical proofs available in the literature for this swap operation were very abstract and we developed our own formal reasoning. In addition, to the best of our knowledge, this is the first formal work on the formalization of a one-dimensional heat equation and the verification of its infinite series solution.

\section{Conclusion}
\label{section6}
In this paper, we proposed a HOL theorem proving based approach for formally analyzing the one-dimensional heat conduction in a rectangular slab. We formalized the heat equation and formally verified its linearity and scaling properties. Moreover, we used the separation of variables method for formally verifying the solution of the heat equation incorporating the corresponding boundary and initial conditions. Next, we formally verified convergence of the generalized solution of the heat equation. Finally, we verified some interesting properties regarding the derivatives of the generalized solution of the heat equation that provide useful insights to the variation of the temperature in the body. In future, we plan to formally verify the uniqueness of the generalized solution of the heat equation and its uniform convergence. Another future direction is to formally analyze the heat transfer in composite slabs \cite{composite}, thermal protection systems \cite{analytical} and heat transfer through various thermoelectric devices, such as thermoelectric generator and thermocouple \cite{thermo} that are widely used in many safety-critical systems.

%

\bibliographystyle{splncs03_unsrt}
\bibliography{heat_conduction}
\end{document}